\documentclass[reprint, superscriptaddress, secnumarabic, amssymb, nobibnotes, aps, prl]{revtex4-1}

\usepackage{graphicx}
\usepackage{epstopdf}
\usepackage[T1]{fontenc}
\usepackage[latin9]{inputenc}
\usepackage{amsbsy}
\usepackage{gensymb}
\usepackage[T1]{fontenc}
\usepackage[latin9]{inputenc}
\usepackage{amsmath}
\usepackage{amssymb}
\usepackage{bbm}
\usepackage{braket}
\usepackage{xcolor}
\allowdisplaybreaks
\usepackage{graphicx}
\usepackage[colorlinks=true]{hyperref}  
\hypersetup{
    bookmarks=true,         
    unicode=false,          
    pdftoolbar=true,        
    pdfmenubar=true,        
    pdffitwindow=false,     
    pdfstartview={FitH},    
    pdftitle={1T-NbSeTe},    
    pdfauthor={a},     
    pdfsubject={},   
    pdfcreator={},   
    pdfproducer={}, 
    pdfkeywords={} {} {}, 
    pdfnewwindow=true,      
    colorlinks=true,       
    linkcolor=blue, 
    citecolor=blue,        
    filecolor=magenta,      
    urlcolor=blue           
} 
\usepackage[normalem]{ulem}

\newcommand{\equref}[1]{Eq.~(\ref{#1})}

\newcommand{\figref}[1]{Fig.~\ref{#1}}

\newcommand{\tableref}[1]{Table~\ref{#1}}

\begin{document}
\title{\textrm{Planar Hall effect and quasi - 2D anisotropic superconductivity in topological candidate 1$T$-NbSeTe}}
\author{C. Patra}
\affiliation{Department of Physics, Indian Institute of Science Education and Research Bhopal, Bhopal, 462066, India}
\author{T. Agarwal}
\affiliation{Department of Physics, Indian Institute of Science Education and Research Bhopal, Bhopal, 462066, India}
\author{Rajeshwari R. Chowdhury}
\affiliation{Department of Physics, Indian Institute of Science Education and Research Bhopal, Bhopal, 462066, India}
\author{R. P. Singh}
\email[]{rpsingh@iiserb.ac.in}
\affiliation{Department of Physics, Indian Institute of Science Education and Research Bhopal, Bhopal, 462066, India}

\begin{abstract}
\begin{flushleft}

\end{flushleft}
Superconducting topological materials have generated considerable interest in condensed matter research due to their unusual gap structures and topological properties. In this study, we have investigated the normal and superconducting characteristics of a potential topological semimetal 1$T$-NbSeTe through comprehensive transport and magnetization measurements on bulk single crystals. The results suggest the topological semimetallic nature of NbSeTe, evidenced by the observation of the planar Hall effect. Moreover, it displays quasi-2D anisotropic superconductivity, which breaks the Pauli limit. The coexistence of the topological semimetallic nature and superconductivity in NbSeTe makes it a potential contender for topological superconductivity.
\end{abstract}
\maketitle

\section{INTRODUCTION}
Superconductors and topological materials are distinct types of quantum materials, both displaying exceptional physical properties with potential applications. When combined, they can produce a distinct quantum phase called a topological superconductor (TSC) \cite{topology_computation,sc_proximity_majarana_Fermi,Majarana_fermion}, representing a new paradigm for unconventional superconductivity. The TSC offers a novel platform to realize non-Abelian quasiparticle excitation, such as Majorana fermions, which can be used as a topological qubit for quantum computation \cite{majarana_quantization,review_TSupercon}. However, the realization of TSC in bulk topological materials is limited. Superconducting Dirac and Weyl semimetals have recently gained significant attention due to their non-trivial topology, as they provide a new platform to innovate TSCs \cite{NiS2_mott_insulator, TSC_chinese, Dirac_Cd3As2, TaAs_weyl,theory_TSC_TaAs}. Meanwhile, doped semimetallic topological transition metal dichalcogenides (TMDs) have emerged as a new platform to realize topological superconductivity \cite{Theory_MoTe2}.\\
 
Recent studies have revealed that the 1$T$ phases of MTe$_2$ (M = Zr, Hf, Ni, Pt, Pd, Ir) \cite{ZrTe2_dirac_semimetal, HfTe2_dirac_semimetal,NiTe2_Re_doped,PtTe2_dirac_semi,Topology_Pd_semi,Pt_doped_sc_IrTe2}, as well as the distorted 1$T''$ phase of NbTe$_2$ \cite{NbTe2_quantum_MR}, exhibit topological semimetallic behavior \cite{band_NbTe2_TaTe2}. Further research suggests that superconductivity can be induced in these materials through strain, pressure, and doping \cite{Pt_doped_sc_IrTe2}, highlighting the potential for 1$T$ phase materials as a platform for topological superconductivity. NbTe$_2$ has garnered significant interest due to its fascinating properties, such as magnetism, charge-density waves (CDW), superconductivity, and quantum spin Hall effect \cite{band_NbTe2_TaTe2,NbTe2_fermi_surface,1T_NbSe2}. Though the 1$T$ phase of NbTe$_2$ and NbSe$_2$ does not occur naturally in bulk crystals, the stable 1$T''$ phase of NbTe$_2$ has been identified as a promising candidate for topological superconductivity, as topological semimetallic behavior and superconductivity ($T_c$ = 0.7 K) coexist in this material \cite{NbTe2_fermi_surface, NbTe2_50T_field}.\\

However, NbSeTe is unique in that it crystallizes in the 1$T$ phase and exhibits enhanced superconductivity due to the suppression of CDW. The disorder is believed to play an important role in both suppressing and enhancing superconductivity in NbSeTe. Recent studies have shown that controlled disorder can enhance the superconducting transition temperature in monolayer NbSe$_2$ and 2$H$-TaSe$_{2-x}$S$_x$ by suppressing competing electronic orders, such as CDW \cite{NbSe2_disorder_effect,effect_of_disorder,tases_2H,TaSe2S2}. Similar disorder-induced enhanced superconductivity has also been observed in bulk 2$H$-TaSe$_{2-x}$S$_x$. These studies suggest that disorder enhanced the electron-phonon coupling strength \cite{angle_resolve_NbSeTe,weak_anti_localization,NbSeTe_reported,TaSeTe_structure,NbTe2_pressur}. Thus, NbSeTe presents a unique opportunity to explore superconductivity in topological semimetals and the impact of the disorder on superconducting topological semimetals.\\

In this paper, we report the synthesis and properties of a single crystal of NbSeTe, which crystallizes in the structure 1$T$ $P\Bar{3}m1$ (164). Our findings reveal that NbSeTe possesses a topological semimetallic nature, as demonstrated by the observation of the planar Hall effect. Furthermore, NbSeTe displays 2D-anisotropic superconductivity with a transition temperature of 3.16 K. The upper critical field surpasses the Pauli limiting field, and its proximity to the unconventional band in the Uemura plot suggests unconventional superconductivity.

\begin{figure*}
	\includegraphics[width=2.08\columnwidth]{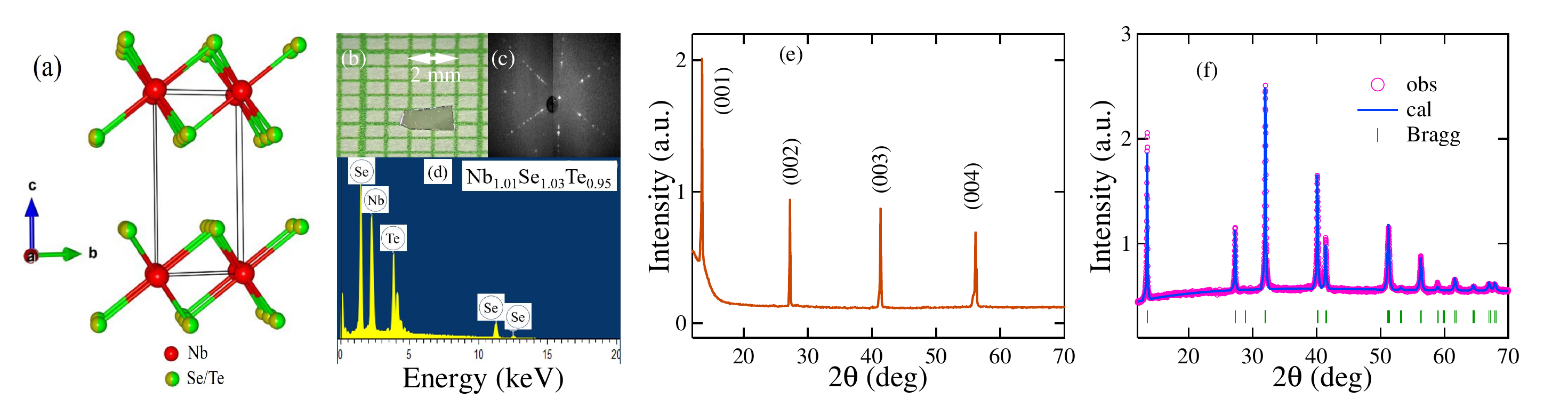}
	\caption{\label{XRD} (a) The unit cell structure of 1$T$-NbSeTe. (b) Shows the image of the single crystal. (c) The Laue diffraction pattern. (d) EDS scan that detects Nb, Se, and Te. (e) The X-ray diffraction pattern of the single crystal indicates that the crystal is oriented along the [00$l$] plane. (f) Refinement of powder XRD data for determining lattice parameters.}
\end{figure*}

\section{EXPERIMENTAL METHODS}

We prepared a single crystal of 1$T$-NbSeTe using the chemical vapor transport method. We mixed Nb (4N), Se (5N), and Te (5N) powders in a stoichiometric ratio and sealed them in a quartz tube with iodine (5 mg cm$^{-3}$) as the transport agent. The tube was then placed in a two-zone horizontal furnace with a temperature gradient of 950\degree C and 850\degree C. After 15 days; shiny single crystals were grown in the cold zone of the tube. The orientation of the crystal plane was determined by X-ray diffraction (XRD) using a PANalytical diffractometer equipped with Cu$K_{\alpha}$ radiation ($\lambda$ = 1.54056 $\text{\AA}$), and the Laue diffraction pattern was recorded using a Photonic-Science Laue camera. We used a scanning electron microscope (SEM) with an energy-dispersive X-ray (EDS) spectrometer to verify the sample compositions, which confirmed the presence of Nb, Se, and Te. Magnetization measurements were conducted using a Quantum Design magnetic measurement system (MPMS3), while resistivity and specific heat measurements were taken using a physical property measurement system (PPMS).

\section{RESULTS AND DISCUSSION}

\subsection{Sample characterization}

\figref{XRD} shows the unit cell, single crystal image, Laue diffraction pattern, EDS measurements, and XRD patterns of 1$T$-NbSeTe. The unit cell, single-crystal image, and Laue diffraction pattern are depicted in (a), (b), and (c), respectively. EDS measurements in various areas of single crystals found the average elemental concentration to be Nb$_{1.01}$Se$_{1.03}$Te$_{0.95}$, as shown in (d). The XRD pattern of the single crystal, depicted in (e), confirms the crystal's orientation along the (00$l$) direction. To determine the phase, we collected the XRD pattern of the crushed single-crystal powder, shown in (f). The powder XRD refinement confirms that the sample crystallizes in the trigonal $P\Bar{3}m1$ (164) space group, with lattice parameters of $a=b=3.56(5)$ \text{\AA} and $c=6.52(7)$ \text{\AA} \cite{NbSeTe_reported,TaSeTe_structure,NbTe2_pressur}. These lattice parameters are identical to those reported for 1$T$-NbSeTe \cite{NbSeTe_reported}.

\subsection{Planar Hall effect}

The planar Hall effect (PHE) is a phenomenon exhibited by topological semimetals (Dirac/Weyl) due to the chiral anomaly and non-trivial Berry curvature of these materials. PHE has been observed in several topological semimetals, including WTe$_2$ \cite{topology_hall_wte}, Cd$_3$As$_2$ \cite{Dirac_Cd3As2}, ZrTe$_5$ \cite{ZrTe5_planar}, PdTe$_2$ \cite{Topology_Pd_semi}, PtTe$_2$ \cite{PtTe2_dirac_semi}, and MoTe$_2$ \cite{origin_planar_hall_MoTe2} as well as superconducting topological materials \cite{LaOBiSe2_planar,TaSe3_planar,Giant_planar_srbi2se3,moirte2}. 
However, recent research suggests that PHE can also be observed in topological materials that lack the chiral anomaly \cite{planr_hall_TI}. Therefore, PHE is considered a useful tool for studying topological behavior through magneto-transport studies.\\

\begin{figure}
\includegraphics[width=1.02\columnwidth]{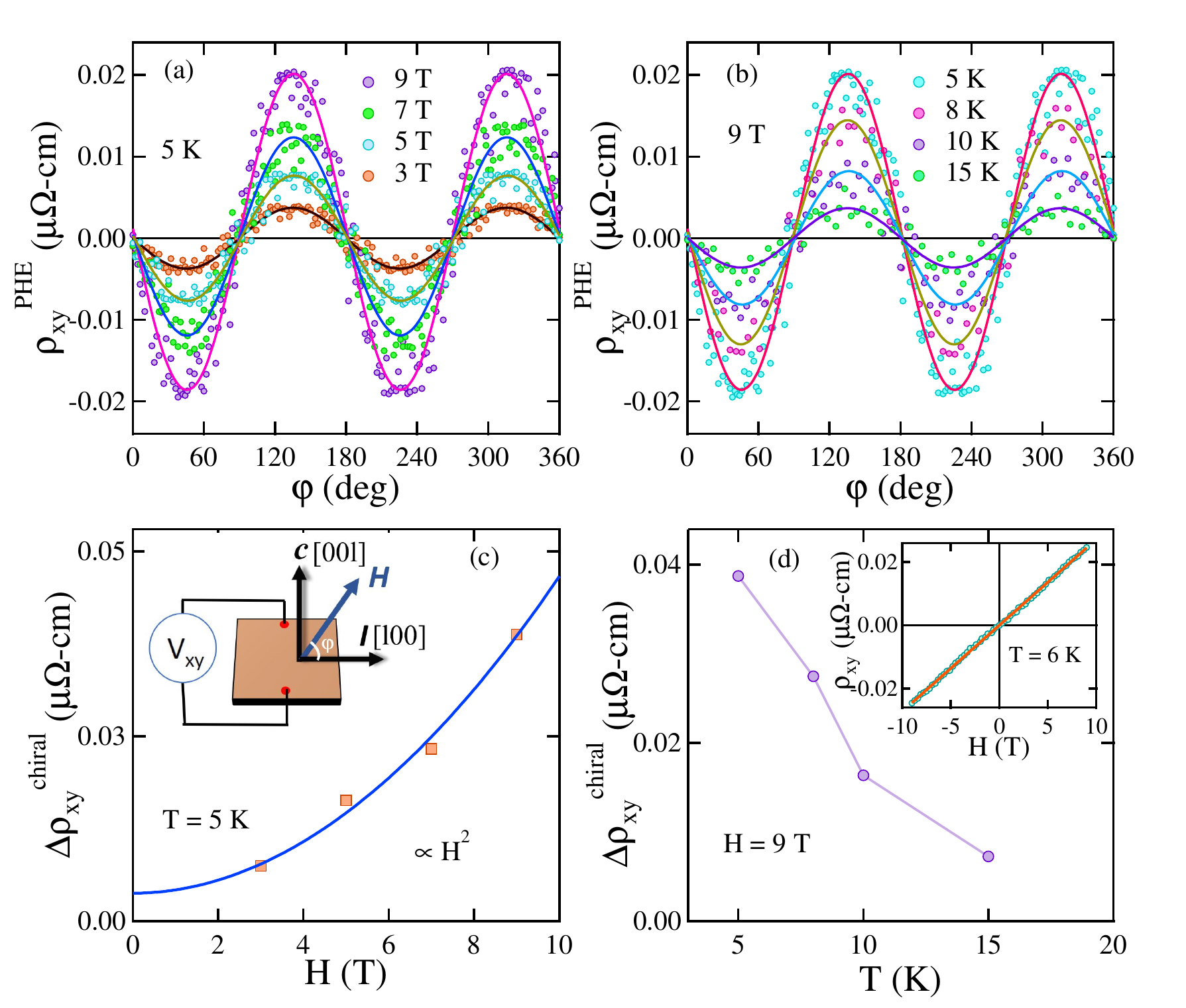}
	\caption{\label{planar_hall}  (a) The planar Hall response for various applied magnetic fields at a temperature of 5 K. (b) Planar Hall effect at different temperatures under 9 T. (c) The variation $\Delta\rho_{xy}^{chiral}$ with the magnetic field. (d) The decrease of $\Delta\rho_{xy}^{chiral}$ with increasing temperature, and an inset showing the Hall resistivity variation with the magnetic field at a temperature of 6 K.}
\end{figure}

The schematic diagram of PHE is represented in the inset of \figref{planar_hall} (c). The standard four-probe technique measures the planar Hall resistivity ($\rho_{xy}$). The magnetic field is applied along the $ab$ plane or perpendicular to the $c$ axis and rotated around the $c$ axis at an angle $\phi$ relative to the current $I$. Experimental Hall resistivity data include normal and planar Hall contributions, with chiral anomaly-induced Hall resistivity symmetric under an antisymmetric applied field, and the normal Hall resistivity being antisymmetric under an opposite applied field \cite{moirte2,planar_hall_review,Cu_doped_PdTe2,mote2_basic_paper}. To extract the chiral anomaly contribution, the planar Hall resistivity data is averaged opposite to the magnetic field as follows:  

\begin{equation}\label{eqn25}
    \rho_{xy}^{PHE} = \left[\frac{\rho_{xy}(B) + \rho_{xy}(-B)}{2}\right]
\end{equation}

where $\rho_{xy}(B)$ and $\rho_{xy}(-B)$ are Hall resistivity under applied positive and negative magnetic field directions, respectively. The angular dependence of $\rho_{xy}^{PHE}$ is shown in \figref{planar_hall} (a) and (b), with a period of $\pi$ and maximum values at $\pi/4$ and $3\pi/4$, consistent with the Planar Hall effect (PHE). In \figref{planar_hall} (a), the variation of planar Hall resistivity is shown at 5 K under different magnetic fields. The variation of $\Delta\rho_{xy}^{chiral}$ with magnetic field strength is proportional to $H^2$, as shown in \figref{planar_hall} (c), indicating the possible presence of a topological phase in the 1$T$-NbSeTe system, similar to NbTe$_2$ \cite{NbTe2_50T_field}. To extract the chiral anomaly contribution, the $\rho{xy}^{PHE}$ data is fitted using \equref{eqn26}. 
\begin{equation}\label{eqn26}
    \rho_{xy} = -\Delta\rho_{xy}^{chiral} sin{\phi}cos{\phi}
\end{equation}

where $\Delta\rho_{xy}^{chiral}$ = $\rho_{\perp} - \rho_{\parallel}$ gives the anisotropy in resistivity due to chiral anamoly. $\rho_{\perp}$ and $\rho_{\parallel}$ represent the resistivity for the magnetic field applied perpendicular and parallel to the current direction, respectively. 
In \figref{planar_hall} (b), the variation of planar Hall resistivity data under 9 T magnetic field at different temperatures is shown. The temperature variation of $\Delta\rho_{xy}^{chiral}$ continuously decreases with increasing temperature, as represented in \figref{planar_hall} (d).\\

Furthermore, the inset of \figref{planar_hall} (d) displays Hall resistivity data ($\rho_{xy}$ vs $H$) fitted with $\rho_{xy} = R_HH$ (where $R_H$ is the Hall coefficient) for 6 K. The obtained carrier concentration is 2.22(3) $\times$ 10$^{21}$ cm$^{-3}$.\\

\subsection{Superconducting properties}

\begin{figure}
	\includegraphics[width=1.05\columnwidth]{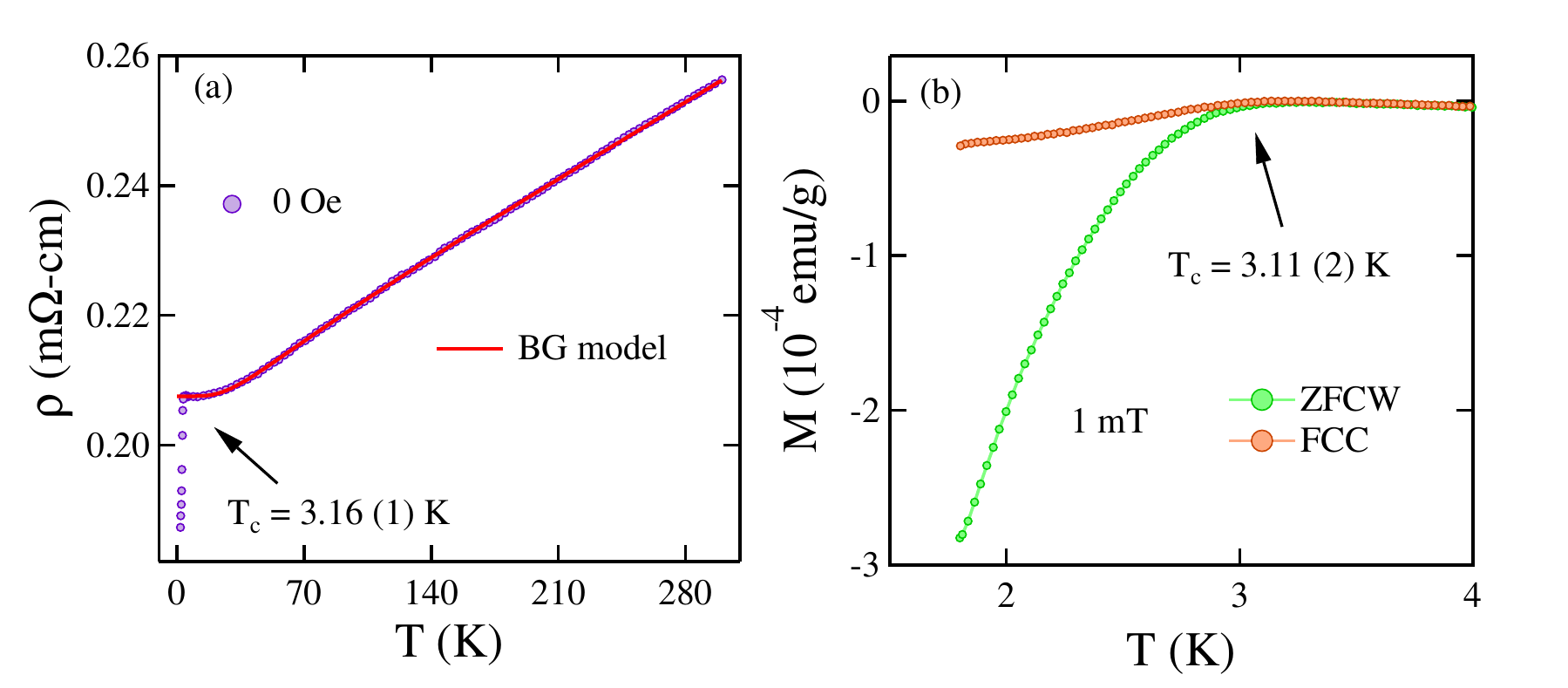}
	\caption{\label{tc} (a) Temperature variation of resistivity at zero magnetic field, indicating superconductivity with a transition temperature of 3.16(1) K. (b)  Magnetization data collected in ZFCW-FCC mode, showing the superconducting transition temperature of 3.11(2) K at an applied magnetic field of 1 mT.}
\end{figure}

Resistivity and magnetization measurements were performed on cleaved crystals (2 mm $\times$ 2 mm). The electrical resistivity vs temperature measurement $\rho(T)$ was conducted over the temperature range of 300 to 1.9 K in the absence of an applied magnetic field, as depicted in \figref{tc} (a). The resistivity shows metallic behavior in the normal state and exhibits a sudden drop at $T_c^{onset}$ of 3.16(1) K due to the superconducting transition, as shown in \figref{tc} (a). The residual resistivity ratio (RRR) is 1.23. In the normal state, the behavior of electrons is influenced by phonons, leading to temperature-dependent behavior. The Bloch-Gr\"{u}neisen (BG) model \cite{BG_model} can explain the temperature dependence of $\rho(T)$ well. According to this model, the resistivity is given by

\begin{equation} \label{eqn11}
\rho (T) = \rho_0 + \rho_{BG} (T)
\end{equation}

where $\rho_{BG}(T)$ is defined as

\begin{equation} \label{eqn12}
\rho_{BG}(T) = r\left(\frac{T}{\Theta_D}\right)^5 \int_{0}^\frac{\Theta_D}{T} \frac{x^5}{(e^x-1)(1-e^{-x})}dx
\end{equation}

here, $\rho_0$ represents the residual resistivity, $r$ is a material-dependent constant, and $\Theta_D$ is the Debye constant. The fitting parameters obtain using \equref{eqn12} were $\rho_0$ = 0.20(7) m$\ohm$ cm, $r$ = 0.11(2) m$\Omega$ cm, and $\Theta_D$ = 169(4) K, as shown in \figref{tc} (a). The Debye temperature value is found to be similar to the reported value in specific heat for NbSeTe \cite{NbSeTe_reported}.\\

The absence of a charge density wave (CDW) signal in the normal state resistivity of 1$T$-NbSeTe distinguishes it from its parent compounds, 1$T''$-NbTe$_2$ and 1$T$-NbSe$_2$, which exhibit strong CDW even in the monolayer limit \cite{cdw_nbte2,1T_NbSe2}. The substitution of selenium in the parent compound may have suppressed the CDW and promoted superconductivity, as reported transition temperatures of 1.3 K \cite{NbSeTe_reported} and ~3.0 K \cite{angle_resolve_NbSeTe,weak_anti_localization} have been observed in 1$T$-NbSeTe without CDW \cite{NbSeTe_reported,angle_resolve_NbSeTe,weak_anti_localization}. In our crystal, the recorded superconducting transition temperature is $T_c$ of 3.16 K, which is comparable to the value reported in Ref. \cite{weak_anti_localization}, revealing that $T_c$ is dependent on the composition.\\

\begin{figure}
	\includegraphics[width=1.02\columnwidth]{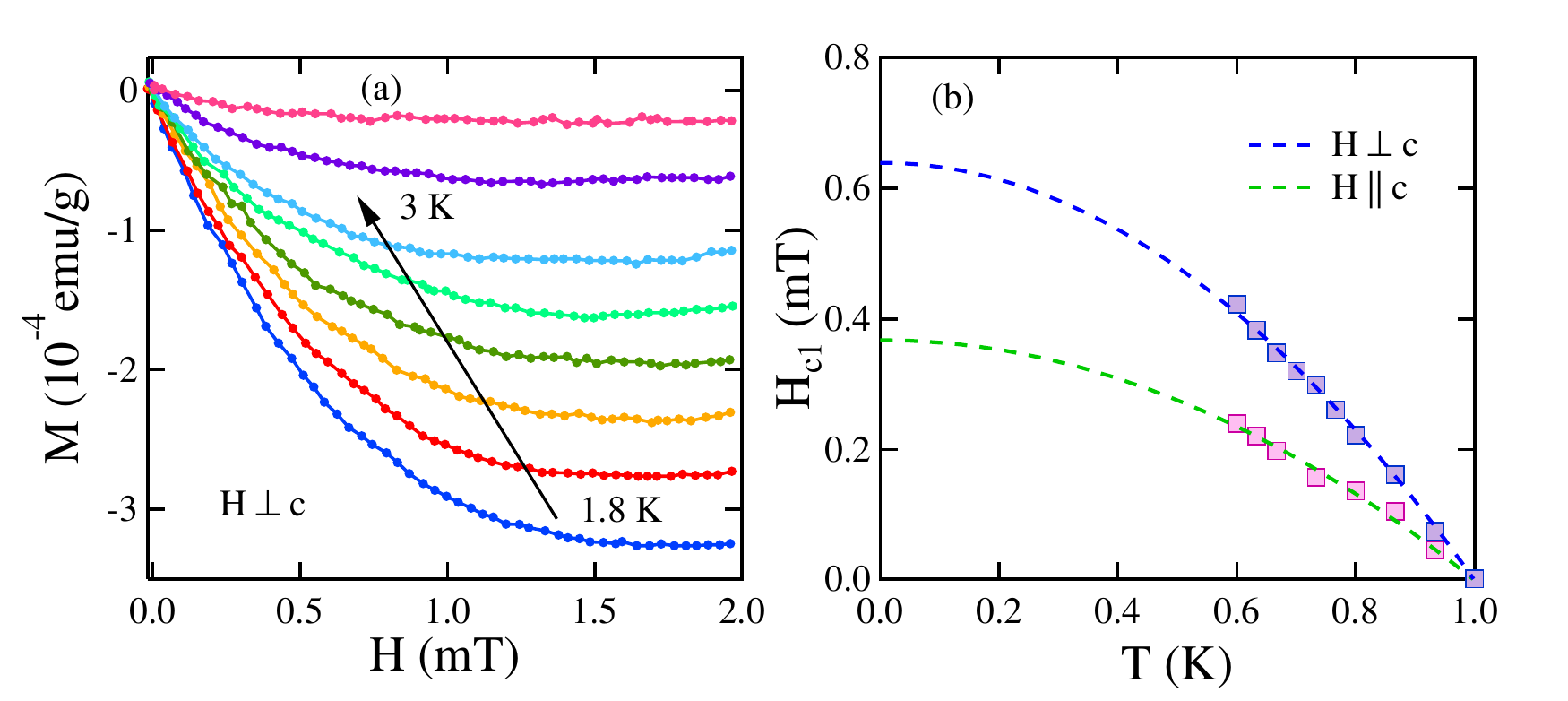}
	\caption{\label{hc10} (a) Low field variation of magnetization data in the perpendicular direction. (b) The lower critical field variation with temperature is well-fitted with GL equations, providing values of 0.63(2) and 0.37(4) mT for the $H \perp c$ and $H \parallel c$ directions, respectively.}
\end{figure}

The temperature dependence of the magnetic moment measurement of 1$T$-NbSeTe with a superconducting transition at 3.11(2) K is presented in \figref{tc} (b). A 1 mT magnetic field was applied in zero-field cooling (ZFCW) and field-cooled cooling (FCC) modes. The lower critical field $H_{c1}$(0) is calculated from low-field magnetization measurements for the perpendicular and parallel directions. The $M$-$H$ curves at various temperatures for $H \perp c$ are illustrated in \figref{hc10} (a). The value of $H_{c1}(T)$ is obtained from the point where the Meissner lines deviate for a specific temperature. The Ginzburg-Landau equation \equref{eqn13} is utilized to fit the values of $H_{c1}(T)$ for both perpendicular and parallel directions of the magnetic field.

\begin{equation} \label{eqn13}
    H_{c1}(T) = H_{c1}(0)\left(1-\left(\frac{T}{T_c}\right)^2\right)
\end{equation}

In \figref{hc10} (b), the anisotropy of $H_{c1}(T)$ in two directions is presented. The blue and green dotted lines correspond to the values of $H_{c1}(0)$, which are 0.63 (2) mT and 0.37(4) mT for the $H \perp c$ and $H \parallel c$ directions, respectively.\\

\begin{figure}
\includegraphics[width=1.02\columnwidth]{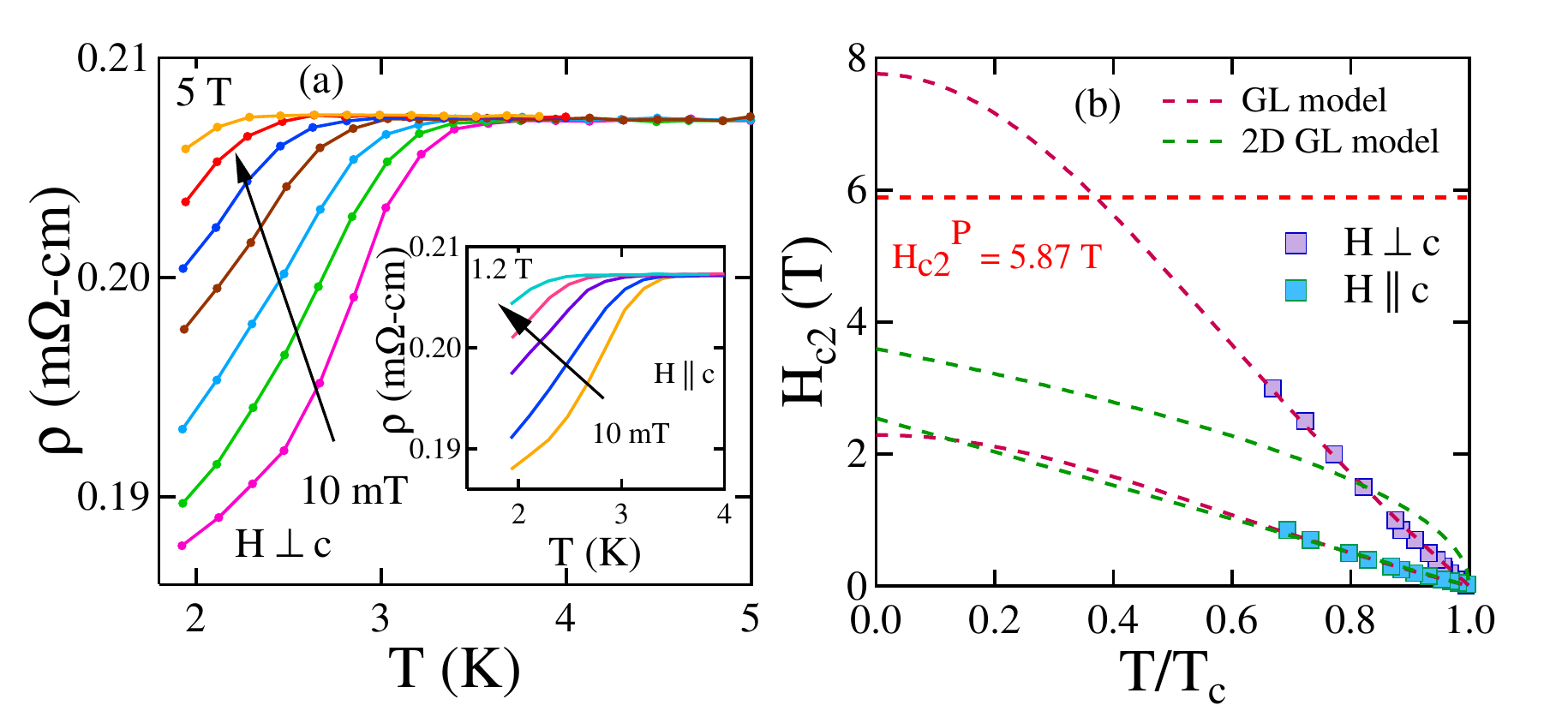}
	\caption{\label{Hc2} (a) show the temperature variation of resistivity for different magnetic fields for H $\perp$ c and H $\parallel$ c in the inset. (b) The temperature variation of the upper critical field for two different magnetic field directions fitted with the Ginzburg-Landau equations. The red dotted line indicates the Pauli limit, H$_{c2}^P$, which is 5.87 T.}
\end{figure}

The upper critical field, $H_{c2}(0)$, was calculated by measuring the isofield resistivity ($\rho$ vs $T$) in the field range of 0 to 5 T, in both the in-plane and out-of-plane directions of the crystal. $H_{c2}(T)$ values are determined from the $96\%$ drop of the normal state resistivity attributed to disorder. \figref{Hc2} (a) displays $\rho$ vs $T$ for the perpendicular and parallel (inset) directions of the magnetic field. The $H_{c2}(T)$ data is fitted using the Ginzburg-Landau equation:

\begin{equation}\label{eqn14}
H_{c2}(T) = H_{c2}(0) \left[\frac{1-\left(\frac{T}{T_c}\right)^2}{1+\left(\frac{T}{T_c}\right)^2}\right].
\end{equation}

The resulting fits in both directions yield the values of 7.76 (1) T and 2.29 (2) T for the $H \perp c$ and $H \parallel c$ direction, respectively, as shown in \figref{Hc2} (b). The 2D Ginzburg-Landau model is used to explain $H_{c2}(T)$, with $(1-T/T_c)^{1/2}$ for $H \perp c$ and $(1-T/T_c)$ for $H \parallel c$. However, the $H_{c2}$ data points deviated for $H \perp c$, whereas for $H \parallel c$ direction, it is consistent with the 3D model, indicating that quasi-2D superconductivity may be present in the system \cite{ausn4}.\\

In type-II superconductors, Cooper pairs can be broken by the orbital effect and the Pauli paramagnetic effect. In the case of the orbital limiting effect, the increasing kinetic energy of the system destroys Cooper pairs. The orbital limiting field ($H_{c2}^{orbital}$) can be expressed using the Wartherm-Helfand-Hohenberg (WHH) equation \cite{paramagnetic_limit}:

\begin{equation}\label{eqn15}
H_{c2}^{orbital} = -\alpha T_c \frac{dH_{c2}(T)}{dT}\bigg|_{T=T_c},
\end{equation}

where $\alpha$ is taken as 0.693 for dirty limit superconductors. For 1$T$-NbSeTe, the initial slope $-\frac{dH_{c2}(T)}{dT}$ near $T_c$ is calculated to be 2.12(1) T/K using the temperature variation of in-plane upper critical field data, giving an orbital limiting field of 4.64(4) T. However, in the Pauli paramagnetic limit, it is energetically favorable for electron spins to align with the magnetic field, breaking the Cooper pairs. The Pauli paramagnetic limit is expressed as $H_{c2}^{P} = 1.86T_c = 5.87(7)$ T (shown as the red dotted line in \figref{Hc2} (b)). Therefore, the upper critical field is 1.32 times higher than the Pauli limiting field. The upper critical fields of comparative materials are listed in \tableref{hc_parameter}. In layered superconductors, violations of the Pauli limit can occur due to strong SOC or finite-momentum pairing. Strong SOC leads to Ising-type superconductivity, which has recently been observed in monolayer 2$H$-NbSe$_2$ \cite{monolayer_nbse2}. Finite-momentum pairing can lead to the Fulde-Ferrell-Larkin-Ovchinnikov (FFLO) state \cite{ba6nb11se28}. More low-temperature angle-dependent measurements and theoretical inputs are required to confirm the exact mechanism of Pauli limit violation.\\

\begin{figure}
\includegraphics[width=1.01\columnwidth]{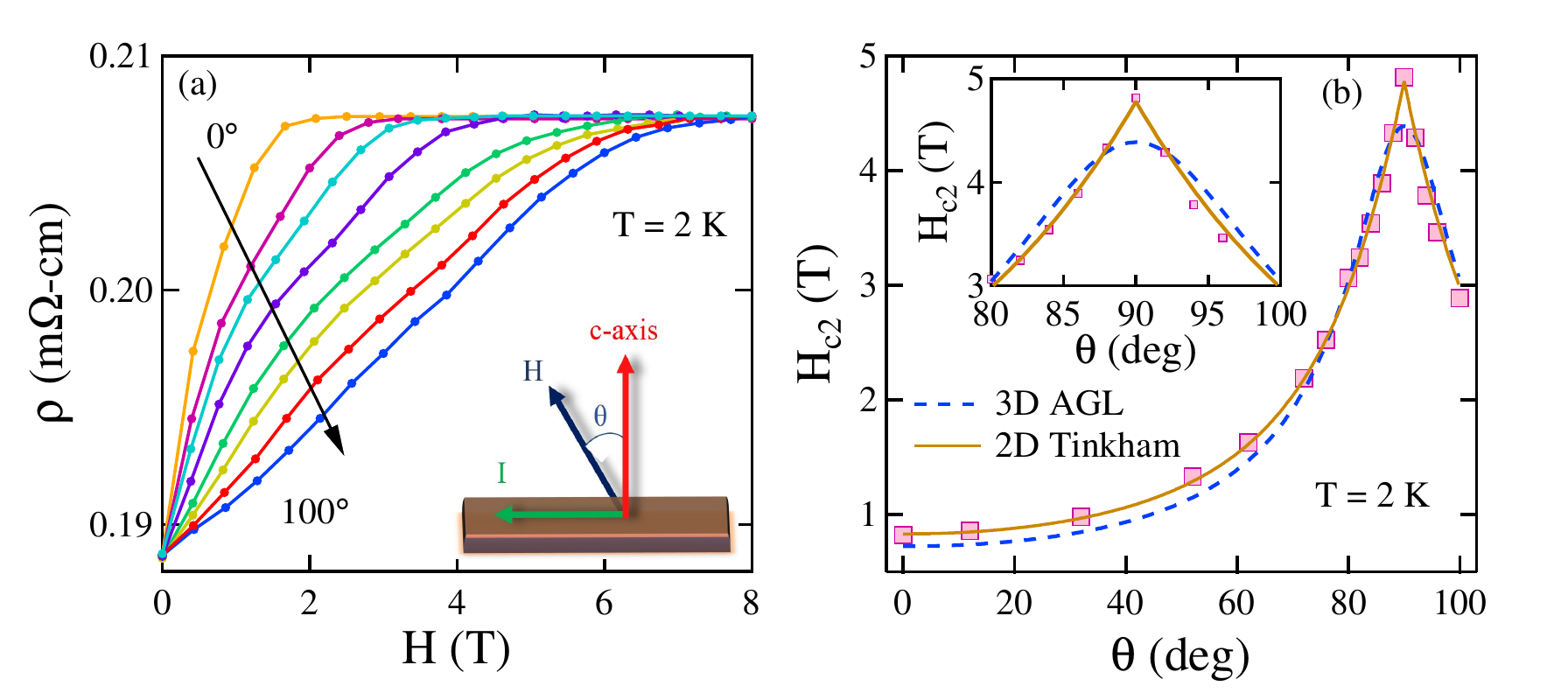}
	\caption{\label{Hc2_angle} (a) The field variation of resistivity at a temperature of 2 K. (b) The angle-dependent upper critical field, fitted with both the 3D Ginzburg-Landau model and the 2D Tinkham model.}
\end{figure}

In addition, a detailed study of the anisotropy in the upper critical field for $H \perp c$ and $H \parallel c$ was conducted by measuring the field-dependent resistivity at 2 K for different angles. The variation of field-dependent resistivity with the angle for 1$T$-NbSeTe is shown in \figref{Hc2_angle} (a), where $\theta$ represents the angle between the magnetic field and the normal to the sample plane (as depicted in the inset of \figref{Hc2_angle} (a)). The $H_{c2}(\theta,T)$ values are obtained from the $98\%$ resistivity drop from the normal state value \cite{ceir3,ausn4}. The anisotropic behavior of the cusp-like upper critical field variation is described using both the 3D anisotropic Ginzburg-Landau model (AGL) and the 2D Tinkham model. 

\begin{equation}\label{eqn16}
    \left(\frac{H_{c2}(\theta,T) \sin{\theta}}{H_{c2}^{\perp}}\right)^2 + \left(\frac{H_{c2}( \theta,T) \cos{\theta}}{H_{c2}^{\parallel}}\right)^2 = 1
\end{equation}

\begin{equation}\label{eqn55}
    \left(\frac{H_{c2}(\theta,T) sin \theta}{H_{c2}^{\perp}}\right)^2 + \left\arrowvert\frac{H_{c2}( \theta,T) cos\theta}{H_{c2}^{||}}\right\arrowvert = 1
\end{equation}

In \figref{Hc2_angle} (b), the field-dependent resistivity is fitted using the 3D anisotropic Ginzburg-Landau model (AGL) and the 2D Tinkham model, represented by the dotted blue and solid green lines, respectively. The zoomed part in the inset of \figref{Hc2_angle} (b) shows that the 2D Tinkham model provides a better fit, indicating the possible presence of 2D superconductivity in the bulk single crystal 1$T$-NbSeTe \cite{ausn4,nbse2_3d_gl}.\\

The Ginzburg-Landau coherence length ($\xi_{\perp c}(0)$ = 119.9(6) $\text{\AA}$ and $\xi_{\parallel c}(0)$ = 35.4(5) $\text{\AA}$) are obtained through the equations $H_{c2}^{\parallel}(0) = \frac{\phi_0}{2\pi\xi_{\perp c}^2(0)} $ and $ H_{c2}^{\perp}(0) = \frac{\phi_0}{2\pi\xi_{\parallel c}(0)\xi_{\perp c}(0)}$ \cite{zeta_calculation}. Utilizing these coherence lengths and the lower critical field values ($H_{c1}^{\perp}(0)$ = 0.63(2) mT and $H_{c1}^{\parallel}(0)$ = 0.37(4) mT), the Ginzburg-Landau penetration depth is calculated to be $\lambda_{\perp c}(0)$ = 1482(9) nm and $\lambda_{\parallel c}(0)$ = 937(7) nm by applying \equref{eqn17}.

\begin{equation}\label{eqn17}
    H_{c1}^{\parallel}(0) = \frac{\phi_0}{4 \pi \lambda^2_{\perp c}(0)}\left[ln\left(\frac{\lambda_{\perp c}(0)}{\xi_{\perp c}(0)}\right) + 0.12\right]
\end{equation}

\begin{table}[h!]
\caption{The comparative superconducting parameters for 1$T$-NbSeTe with some layered compounds}
\label{hc_parameter}
\begin{center}
\begin{tabular*}{1.0\columnwidth}{l@{\extracolsep{\fill}}cccc}\hline\hline
Parameter& 1$T$- & 2$H$- & NbS$_2$ & Ba$_6$Nb$_{11}$Se$_{28}$ \\
& NbSeTe & NbSe$_2$\cite{pauli_NbSe2} & \cite{pauli_2hNbSe2} & \cite{ba6nb11se28}\\
\hline
\\[0.5ex]
$H_{c2}^{\perp}$(T) & 7.76 & 17.3 & &8.84\\
$H_{c2}^{\parallel}$(T) & 2.29 & 5.3 & 1.6 & 0.57\\
$H_{c2}^{P}$(T) & 5.87 & 13.54 & 10.4 & 4.27\\
$\xi_{\perp c}(\text{\AA}$) & 119.9 & 78.8 & 143 & 240.4\\
$\xi_{\parallel c}(\text{\AA}$) & 35.4 & 24.1 & 9.7 & 15.6\\
$\gamma =\frac{H_{c2}^{\perp}(0)}{H_{c2}^{\parallel}(0)}$ & 3.39 & 3.3 & 7.94 & 10.53\\
\\[0.5ex]
\hline\hline
\end{tabular*}
\par\medskip\footnotesize
\end{center}
\end{table}

where $\phi_0$ (= 2.07$\times$10$^{-15}$ Tm$^2$) is the quantum magnetic flux. Using the two coherence lengths, the Ginzburg-Landau parameter $\kappa_{\perp c}$ is calculated as 123$>$ $\frac{1}{\sqrt{2}}$ by $\kappa_{\perp c} = \frac{\lambda_{\perp c}(0)}{\xi_{\perp c}(0)}$. This suggests that the sample exhibits type II superconducting behavior.\\

The electron-phonon coupling constant ($\lambda_{e-ph}$) is a dimensionless parameter that provides insight into the strength of interactions within the system. Using McMillan's model \cite{coupling_electron_phono}, $\lambda_{e-ph}$ can be calculated from $\Theta_D$ and $T_c$ using 169 K and 3.16 K respectively in \equref{eqn141}, where $\mu^*$ is the Coulomb repulsion constant (0.13 in our case).

\begin{equation}\label{eqn141}
\lambda_{e-ph} = \frac{1.04+\mu^*\ln(\Theta_D/1.45T_c)}{(1-0.62\mu^*)\ln(\Theta_D/1.45T_c)-1.04}
\end{equation}

For our sample, the calculated value of $\lambda_{e-ph}$ is 0.66, which is higher than the reported value of 0.55 for 1$T$-NbSeTe with $T_c$ of 1.3 K. This enhancement in $\lambda_{e-ph}$ suggests that the weak disorder can increase the electron-phonon coupling and consequently enhance the superconducting transition temperature. This has been observed in other systems, such as weak disorder in 2$H$-TaSe$_{2-x}$S$_x$ \cite{TaSe2S2}, Ir-doped MoTe$_2$ \cite{moirte2}, and Re-doped NiTe$_2$ \cite{NiTe2_Re_doped}.\\

\subsection{Electronic properties and the Uemura plot}

A set of equations are solved to determine the electronic mean free path and to verify the dirty or clean limit of the 1$T$-NbSeTe superconductor \cite{lamda_zeta_calculation}. The Sommerfeld coefficient, which is related to the quasiparticle number density per volume with mean free path, is given by \equref{eqn22}

\begin{equation}\label{eqn22}
    \gamma_n = \left(\frac{\pi}{3}\right)^{2/3}\frac{k_B^2m^*V_{f.u.}n^{1/3}}{\hbar^2N_A},
\end{equation}

where $k_B$ is the Bolzman constant, $m^*$ is effective mass, $V_{f.u.}$ is number density per volume of the quasiparticle, and $N_A$ is Avogadro number. The relationship between the mean free path $l_e$, the residual resitivity $\rho_0$, and the Fermi velocity is

\begin{equation}\label{eqn23}
    l_e = \frac{3\pi^2\hbar^3}{e^2\rho_0m^{*2}v_F}.
\end{equation}

The value of the Sommerfeld coefficient used in this study is 5.7 mJ/mol K$^{-2}$, obtained from the previous report on specific heat measurement for 1$T$-NbSeTe \cite{NbSeTe_reported}. Using carrier concentration and residual resistivity, the effective mass, Fermi velocity, and mean free path are calculated to be 6.27 m$_e$, 0.74 $\times$ 10$^{5}$ m/s and 36.05 (6) $\text{\AA}$, respectively, with a coherence length of 324(2) $\text{\AA}$ \cite{lamda_calculation}. The $\frac{\xi_0}{l_e}$ ratio is 9, indicating that the 1$T$-NbSeTe sample is in the dirty limit.\\

\begin{figure}
\includegraphics[width=1.0\columnwidth]{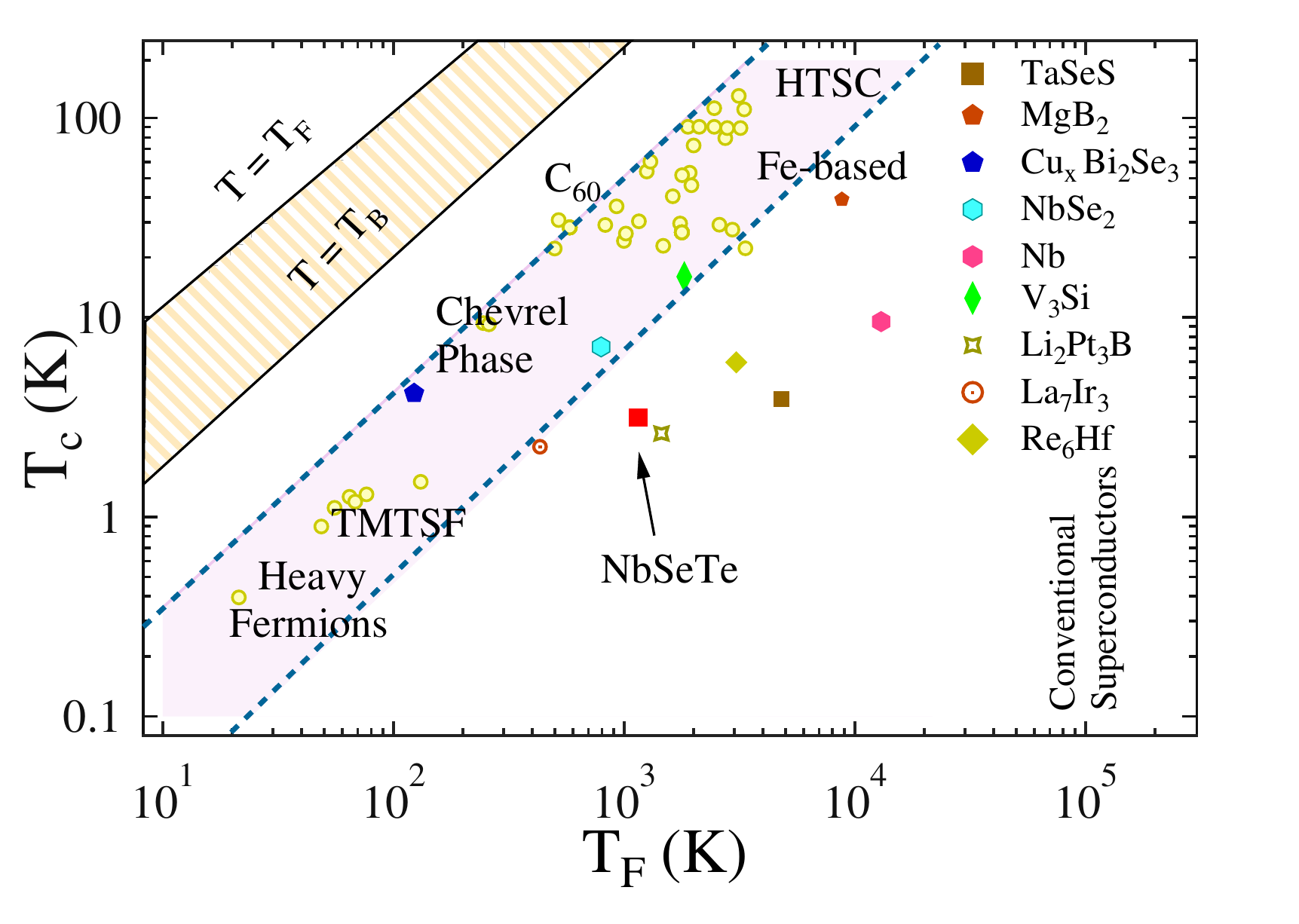}
	\caption{\label{Uemura}  The plot of the superconducting transition temperature vs the Fermi temperature for different superconducting families. In between, two solid blue lines show the unconventional band of superconductors. NbSeTe lies close to the unconventional band.}
\end{figure}

According to Uemura et al., superconducting materials can be classified as conventional or unconventional based on their $T_c/T_F$ ratio. Unconventional superconductors such as Chevrel phases, heavy fermions, Fe-based superconductors, and high $T_c$ superconductors have a $T_c/T_F$ ratio within the range of 0.01 $\le$T$_c$/T${_F}\le$0.1 \cite{uemra_plot_1,umara_plot_2}. To determine the classification of 1$T$-NbSeTe, the Fermi temperature $T_F$ was evaluated using the equation:

\begin{equation}\label{eqn24}
k_BT_F = \frac{\hbar^2}{2} (3\pi^2)^{2/3}\frac{n^{2/3}}{m^*}
\end{equation}

where n is the carrier density of 2.22 $\times$ 10$^{27}$ m$^{-3}$ and m$^*$, the effective mass is 6.27m$_e$ for the sample. The calculated Fermi temperature for NbSeTe is 1151(4) K (as shown in \figref{Uemura}). The obtained $T_c/T_F$ ratio of 0.0027 is in the unconventional range, similar to other unconventional materials, which may indicate the presence of a topological phase.

\section{Conclusion}

In summary, we have successfully synthesized single crystals of NbSeTe in the trigonal 1$T$ phase with space group 164. The material exhibits exotic chiral anomaly-induced PHE with $H^2$ behavior and has a superconducting transition temperature of 3.16 K. The in-plane upper critical field exceeds the Pauli limit, and the angle-dependent upper critical field suggests the presence of 2D superconductivity, indicating unconventional superconductivity in bulk NbSeTe. Furthermore, the Uemura plot also supports the presence of unconventional superconductivity in this material. The coexistence of topological semimetallic nature and superconductivity in NbSeTe makes it a potential candidate for topological superconductivity. However, further investigations, including angle-resolved photoemission spectroscopy (ARPES), detailed electronic structure calculations, and low-temperature and thickness-dependent measurements, are necessary to fully understand this system's superconducting pairing mechanism.

\section{Acknowledgments}

R.~P.~S.\ acknowledge Science and Engineering Research Board, Government of India, for the CRG/2019/001028 Core Research Grant.

\end{document}